\documentstyle[twoside,12pt]{article}

\catcode`\@=11
\long\def\@makefntext#1{ 
\protect\noindent \hbox to 3.2pt {\hskip-.9pt
$^{{\eightrm\@thefnmark}}$\hfil}#1\hfill} 

\def\thefootnote{\fnsymbol{footnote}}
 \def\@makefnmark{\hbox to 0pt{$^{\@thefnmark}$\hss}}  

\def\ps@myheadings{\let\@mkboth\@gobbletwo
\def\@oddhead{\hbox{} 
\rightmark\hfil\eightrm\thepage}
\def\@oddfoot{}\def\@evenhead{\eightrm\thepage\hfil 
\leftmark\hbox{}}\def\@evenfoot{}
\def\sectionmark##1{}\def\subsectionmark##1{}}

\oddsidemargin=\evensidemargin
\renewcommand{\thefootnote}{\fnsymbol{footnote}}

\newcounter{sectionc}\newcounter{subsectionc}\newcounter{subsubsectionc}
\renewcommand{\section}[1] {\vspace{12pt}\addtocounter{sectionc}{1}
\setcounter{subsectionc}{0}\setcounter{subsubsectionc}{0}\noindent
	{\bf\thesectionc. #1}\par\vspace{5pt}}
\renewcommand{\subsection}[1] {\vspace{12pt}\addtocounter{subsectionc}{1}
	\setcounter{subsubsectionc}{0}\noindent
	{\bf\thesectionc.\thesubsectionc. {\kern1pt \bf\it #1}}\par\vspace{5pt}}
\renewcommand{\subsubsection}[1] {\vspace{12pt}\addtocounter{subsubsectionc}{1}
	\noindent{\thesectionc.\thesubsectionc.\thesubsubsectionc.
	{\kern1pt \it #1}}\par\vspace{5pt}}
\newcommand{\nonumsection}[1] {\vspace{12pt}\noindent{\bf #1}
	\par\vspace{5pt}}

\newcommand{\textlineskip}{\baselineskip=14pt}

\def\eightcirc{
\begin{picture}(0,0)
\put(4.4,1.8){\circle{6.5}}
\end{picture}}
\def\eightcopyright{\eightcirc\kern2.7pt\hbox{\eightrm c}}

\def\abstracts#1#2#3{{
	\centering{\begin{minipage}{5in}\baselineskip=12pt\tenrm
	\centerline{ABSTRACT}
	\parindent=0pt #1\par
	\parindent=15pt #2\par
	\parindent=15pt #3
	\end{minipage} }\par}}

\renewenvironment{thebibliography}[1]			
	{
	 \begin{list}{\arabic{enumi}.}			
	{\usecounter{enumi}\setlength{\parsep}{0pt}
	 \setlength{\leftmargin 17pt}{\rightmargin 0pt}	
	 \setlength{\itemsep}{0pt} \settowidth		
	{\labelwidth}{#1.}\sloppy}}{\end{list}}	

\newcounter{itemlistc}
\newcounter{romanlistc}
\newcounter{alphlistc}
\newcounter{arabiclistc}

\newenvironment{romanlist}
	{\setcounter{romanlistc}{0}
	 \begin{list}{(\roman{romanlistc})}
	{\usecounter{romanlistc}
	 \setlength{\parsep}{0pt}
	 \setlength{\itemsep}{0pt}
	}}{\end{list}}

\newcounter{tempfigtabc}			

\def\pmb#1{\setbox0=\hbox{#1}
	\kern-.025em\copy0\kern-\wd0
	\kern.05em\copy0\kern-\wd0
	\kern-.025em\raise.0433em\box0}

\def\fnt#1#2{\footnotetext{\kern-.3em
	{$^{\mbox{\scriptsize #1}}$}{#2}}}

\def\fpage#1{\begingroup
\voffset=.3in
\thispagestyle{empty}\begin{table}[b]\centerline{\footnotesize #1}
	\end{table}\endgroup}

 1
 1
 1

\font\eightrm=cmr8

\def\qed{\hbox{${\vcenter{\vbox{                          
   \hrule height 0.4pt\hbox{\vrule width 0.4pt height 6pt
   \kern5pt\vrule width 0.4pt}\hrule height 0.4pt}}}$}}

\def\@cite#1{\mbox{\small $^{#1}$}}
\begin{document}
\normalsize\textlineskip
{\thispagestyle{empty}
\setcounter{page}{1}

\renewcommand{\thefootnote}{\fnsymbol{footnote}} 
\def\bsc{{\sc a\kern-6.4pt\sc a\kern-6.4pt\sc a}}
\def\bflatex{\bf L\kern-.30em\raise.3ex\hbox{\bsc}\kern-.14em
T\kern-.1667em\lower.7ex\hbox{E}\kern-.125em X}
\newcommand\nn{\nonumber}
\newcommand\beq{\begin{equation}}
\newcommand\eeq{\end{equation}}
\renewcommand\d{\partial}
\newcommand\eps{\epsilon}
\newcommand\fpi{f_\pi}
\newcommand\half{{1\over 2}}
\newcommand\fourth{{1\over 4}}
\newcommand{\dg}{\mbox{$^\dagger$}}
\renewcommand\Im{{\rm Im}}
\renewcommand\Re{{\rm Re}}
\newcommand\be{\begin{eqnarray}}
\newcommand\ee{\end{eqnarray}}
\newcommand{\equa}[1]{(\ref{#1})}
\newcommand{\equas}[2]{(\ref{#1}--\ref{#2})}
\newcommand{\tss}{\textstyle}
\newcommand{\psia}{\psi_1 (\mu)}
\newcommand{\psib}{\psi_2 (\mu)}
\newcommand{\psic}{\psi_3 (\mu)}
\newcommand{\psid}{\psi_4 (\mu)}
\newcommand{\psiz}{\psi (\mu)}
\newcommand{\bmu}{\mbox{$\bar{ \mu }$}}
\newcommand{\Frac}[2]{{\frac{\displaystyle #1}{\displaystyle #2}}}
\newcommand{\cD}{{\cal D}}
\newcommand{\cE}{{\cal E}}
\newcommand{\cG}{{\cal G}}
\newcommand{\cH}{{\cal H}}
\newcommand{\cL}{{\cal L}}
\newcommand{\Tr}{{\rm Tr}}
\newcommand{\phia}{\mbox{$\Phi^{\alpha}$}}
\newcommand{\phib}{\mbox{$\Phi^{\beta}$}}
\newcommand{\dipa}{\partial_i \Phi^a}
\newcommand{\djpa}{\partial_j \Phi^a}
\newcommand{\del}{\partial}
\newcommand{\delphia}{\mbox{$\delta \Phi^{\alpha}$}}
\newcommand{\delphib}{\mbox{$\delta \Phi^{\beta}$}}
\newcommand{\ltoinf}{$L \to \infty$}
\newcommand{\sthree}{\mbox{${\rm \bf S^{3}}(L)$}}
\newcommand{\sunit} {\mbox{${\rm \bf S^{3}}(1)$}}
\newcommand{\rflat}{${\rm \bf {R}^{3}}$}
\newcommand{\otwo}{${\rm O(2)}$}
\newcommand{\lag}[1]{{\cal L}_{#1}}
\newcommand{\prc}[3] {{\em Phys.\,Rev.}{\bf C#1}, #2 (19#3)}
\newcommand{\prd}[3] {{\em Phys.\,Rev.} {\bf D#1}, #2 (19#3)}
\newcommand{\prold}[3]{{\em Phys.\,Rev.} {\bf #1}, #2 (19#3)}
\newcommand{\plb}[3] {{\em Phys.\,Lett.} {\bf #1 B}, #2 (19#3)}
\newcommand{\prl}[3] {{\em Phys.\,Rev.\,Lett.} {\bf #1}, #2 (19#3)}
\newcommand{\np}[3] {{\em Nucl.\,Phys.} {\bf #1}, #2 (19#3)}
\newcommand{\npa}[3] {{\em Nucl.\,Phys.} {\bf A#1}, #2 (19#3)}
\newcommand{\npb}[3] {{\em Nucl.\,Phys.} {\bf B#1}, #2 (19#3)}
\newcommand{\jpa}[3] {{\em J.\,Phys.} {\bf A#1}, #2  (19#3)}
\newcommand{\jpb}[3] {{\em J.\,Phys.} {\bf B#1}, #2  (19#3)}
\newcommand{\jpc}[3] {{\em J.\,Phys.} {\bf C#1}, #2  (19#3)}
\newcommand{\prep}[3] {{\em Phys.\,Rep.} {\bf #1}, #2 (19#3)}
\newcommand{\prog}[3] {{\em Prog.\,Theor.\,Phys.} {\bf #1}, #2 (19#3)}
\newcommand{\commath}[3] {{\em Commun.\,Math.\,Phys.} {\bf #1}, #2 (19#3)}
\newcommand{\lmathphys}[3] {{\em Lett.\,Math.\,Phys.} {\bf #1}, #2 (19#3)}
\newcommand{\advnp}[3] {{\em Adv.\,Nucl.\,Phys.} {\bf #1}, #2 (19#3)}
\newcommand{\reppp}[3] {{\em Rep.\,Prog.\,Phys.} {\bf #1}, #2 (19#3)}
\newcommand{\jetp}[3] {{\em JETP} {\bf #1}, #2 (19#3)}
\newcommand{\jetplett}[3] {{\em JETP Lett.} {\bf #1}, #2 (19#3)}
\newcommand{\ap}[3] {{\em Ann.\,Phys. (N.Y.)} {\bf #1}, #2 (19#3)}
\newcommand{\rmB}{{\rm B}}
\newcommand{\rmL}{{\rm L}}

\fpage{1}
\centerline{\bf NUCLEAR MATTER ASPECTS OF SKYRMIONS\footnote{Talk at
the workshop on ``Baryons as Skyrme Solitons'', Siegen, Germany,
Sept. 27-30, 1992. TH-Darmstadt-Preprint IKDA 92/37 ---
hep-ph/9211295}}
\vspace{0.3truein}
\centerline{\footnotesize ANDREAS WIRZBA}
\vspace*{0.015truein}
\centerline{\footnotesize\it Institut f\"{u}r Kernphysik,
Technische Hochschule Darmstadt, Schlo{\ss}gartenstr.~9}
\baselineskip=12pt
\centerline{\footnotesize\it D-W-6100 Darmstadt, Germany}

\vspace*{0.21truein}
\abstracts{As an alternative approach to the infinite-array description of
dense matter in the Skyrme model, we report about the properties of a
single skyrmion on a compact 3-sphere of finite radius.  The density
of this matter can be increased by decreasing the hypersphere radius.
As in the array calculations one encounters a transition to a distinct
high density phase characterized by a delocalization in energy and
baryon charge and by increased symmetries. We will argue that the high
density phase has to be interpreted as chirally restored one. The
arguments are based on the formation of complete chiral multiplets and
the vanishing of the pionic massless Goldstone modes in the
high-density fluctuation spectrum.  We show that the restoration of
chiral symmetry is common to any chirally invariant extension of the
usual Skyrme model - whether via higher-order contact terms or via the
introduction of stabilizing vector mesons which act over a finite
range.  }{}{}

\vspace*{-3pt}\textlineskip
\section{Introduction}
\textheight=8.5truein
\setcounter{footnote}{0}
\renewcommand{\thefootnote}{\alph{footnote}}
In this lecture we will present some studies about the high-density
behavior of baryonic configurations in the Skyrme model~\cite{Skyrme}
and its variants~\cite{Ulfreport,structure}. These models belong to a
class of effective models which treat the baryon stabilization (and
hence the baryon structure) and the interaction between baryons on the
same footing. There is no principal difference between the
stabilization and the interaction mechanism, the models just act in
different topological sectors: in the baryon (=winding) number $B=1$
sector for the structure physics~\cite{ANW,structure} and in the $B>1$
sector for the interaction physics~\cite{interaction}.  For
$B\to\infty$ one has a new parameter (the density) at ones disposal in
order to tune the interplay between the stabilization and the
interaction mechanisms. Because of these features the Skyrme model and
its variants have the inherent possibility to allow for radical
changes when baryonic matter is compressed to high densities: the
baryons which at low densities are well-separated and clearly defined
objects might completely loose their identity and the baryon matter
can become uniform. There are at least two distinct ways for
investigating the high-density behavior of Skyrme-type models.

The first approach which was pioneered by Igor
Klebanov~\cite{Klebanov} uses infinite periodic arrays of skyrmions
with baryon number (winding number) of one per unit cell in real
space-time. The details of the lattice structure and the periodic
boundary conditions to which the mesonic fields are subject can be
chosen as to avoid nearest neighbor frustration at low densities.  The
aim is to find the classical static field configuration which
minimizes the energy per cell volume for a fixed baryon charge per
array cell. The equations of motion (which are non-linear partial
differential equations) are solved numerically on a lattice grid
subject just to two constraints: the choice of the crystal and the
form of the {\em twisted} periodic boundary conditions which guarantee
the periodic structure and minimize the frustrations in the field
gradients (and thus the energy) between neighboring cells in the
asymptotic low density region. The boundary conditions are then
extrapolated without alteration to high densities as well. At low
densities the skyrmions in the periodic arrays are well-separated,
ensembled in a phase of weakly interacting baryons. With increased
density, however, they grow in size as measured by their r.m.s.\
radius until at a critical density - as first observed by W\"{u}st,
Brown and Jackson~\cite{Wuest} - they ``melt'' to a distinct
high-density phase where the skyrmions completely lose their identity.
The new phase is characterized by the approximate uniformity in the
baryon as well as the energy density, by the fact that the averages of
the $\sigma$- and $\pi$-fields vanish over the cell volume and that
there appears an additional symmetry at the critical density: the
``half-skyrmion symmetry''\,\footnote{This symmetry is characterized by
families of planes on which the scalar field, $\sigma$ is zero. The
pionic fields have a reflection symmetry about these planes while the
scalar field is reflection antisymmetric.} of Goldhaber and
Manton.\cite{Goldhaber}

The short-comings of these calculations are: they are by fiat of
numerical nature, the rotational and translational symmetries are
explicitly broken down to discrete ones by the artificial crystalline
grid structure. Finally the calculations depend on the form of the
crystal and the twisted periodic boundary conditions which are
extracted in the asymptotic low density regime, but nevertheless
applied at high densities. For instance, the order of the phase
transition can depend on these choices: For the same simple cubic
lattice it is first order for Klebanov's choice of boundary terms,
while it is second order for the modified rectangular boundary terms
of Jackson and Verbaarschot~\cite{AndyJac} where a preferred direction
in space is singled out. But the fact that there are phase transitions
are common for all these calculations - and even the critical
densities are approximately the same. Klebanov's crystal is not of
minimal energy, a face-centered cubic arrangement of skyrmions (at low
densities) is preferable. Again, as the density increases there is a
second-order phase transition and the half-skyrmion symmetry
emerges.\cite{fcc}\ The crystal of minimal energy is in this higher
density phase and has energy per baryon only of 3.8\% above the
topological lower bound. Further generalizations are the inclusion of
temperature and the derivation of the equation of state for the Skyrme
matter in the case of the simple cubic lattice~\cite{Walhout} as well
as the fcc structure~\cite{Walhoutfcc} by T.S.~Walhout. He has also
studied periodic arrays using an $\omega$-stabilized variant of the
Skyrme model.\cite{Walhoutomega}

Nevertheless, the question might arise whether these phase transitions
are lattice artifacts or not.  Fortunately, there is an alternative
approach to study the same phenomena by replacing the periodic arrays
in flat space \rflat\ by few-skyrmion-systems (or even a single
skyrmion) on the compact manifold \sthree\ pioneered by Nick
Manton.\cite{Ruback,Manton}\  The finite baryon number on \sthree\
corresponds to a finite baryon density in \rflat (with infinite baryon
number), so that one obtains a model for skyrmion matter which is
appreciably simpler to study than any lattice model. The density of
this matter can be increased by decreasing the hypersphere radius $L$,
the radius of the 3-sphere in four dimensions. In the limit
$L\to\infty$ a skyrmion localized on the hypersphere has basically the
same properties as an isolated skyrmion in flat space or in a periodic
array for infinite separation. For a large hypersphere it is still
well localized and dominated by its stabilization mechanism, but
because of the curvature and finite size effects of the hypersphere it
acts as if the tail of another skyrmion (here of course nothing else
than itself seen via the opposite pole of the 3-sphere) is present. By
shrinking the hypersphere radius this interaction mechanism will
become stronger and stronger. Finally it can become so strong that the
stabilizing mechanism cannot localize the skyrmion any longer: the
skyrmion will be smeared (=delocalized) over the whole 3-sphere.  The
connection between both approaches can be made by identifying the
averaged baryon densities: in the periodic array calculation the
baryon density is averaged over the cell volume, on the hypersphere it
is given by the ratio $B/(2\pi^2 L^3_{\rm dim})$ of the baryon charge
$B$ and the (surface-) volume of the 3-sphere in terms of the
dimensioned radius $L_{\rm dim}$. The \sthree\ approach has the
advantage that it is technically far simpler than the period array
one\,\footnote{In fact most of the calculations can be done
analytically or involve at most the numerical task of solving ordinary
non-linear differential equations.} and that it allows for
mathematically rigorous results for the properties of the
``high-density'' (or ``small $L$'') phase of a single skyrmion on
\sthree. Furthermore neither the (continuous) rotational nor the
translational symmetry are broken.  The disadvantage is obvious: The
physical space is not a hypersphere - at least not a small one. Both
approaches supplement each other in a complementary way: the first is
realized in normal space-time and for infinite systems (a precondition
for a phase transition), but is technically complicated and plagued by
its crystalline nature, by the ambiguities in choosing the array
itself and the boundary terms; the other is simple and exact, but is
set up in an unphysical world and limited to finite systems which
therefore should be not identified with an infinite periodic array,
but just with {\em one} array cell.

The talk is organized as follows. In sect.~2 we will give Manton's
general description of the Skyrme model on the 3-sphere and review
some of the $B=1$ properties. In sect.~3 we present arguments why the
phase transition has to be interpreted as chiral restoration. In
sect.~4 we report on the generalization to variants of the
Skyrme-model which involve higher-order contact terms or which are
stabilized by vector mesons. Sect.~5 contains a discussion about the
order of the phase transition. We end the talk with a short discussion
section.

\section{The Skyrme Model on the Hypersphere}
As mentioned there is considerable interest in studying the
Skyrme model
\beq
  \lag{2,4} = \Frac{{f_\pi}^2}{4}\Tr \left( \partial_\mu U\dg
                                      \partial^\mu U \right )
              + \Frac{{\epsilon_4}^2}{4}
   \Tr \left [ U^\dagger \partial_\mu U , U^\dagger \partial_\nu U
\right ]^2 ,
  \label{L24}
\eeq
on a 3-dimensional hypersphere of radius $L$,
\sthree.\cite{Ruback,Manton}\  We will report here especially on the
findings of refs.\cite{Manton,Loss,S3}.  In order to get
parameterization-independent results it is useful to relate the
quaternion representation $U$ of the Skyrme model to a cartesian
representation $\{\Phi^\alpha\}= (\Phi^0,\Phi^z,\Phi^x,\Phi^y)$ with
the imposed constraint $\Phi^\alpha
\Phi^\alpha =1$ as $U=\Phi^0 + i {\vec\tau}\cdot{\vec \Phi}$ and
to introduce the strain tensor~\cite{Manton}
\beq
 \begin{array}{lcl}
   K_{i j} &\equiv& \dipa \djpa \qquad\qquad
          \left ( {\rm with}\quad \{ \Phi^\alpha \} =( \sigma,
                                                 \pi^z, \pi^x, \pi^y
)\right )\\
           &=& -\frac{1}{4} \Tr \{ U\dg \del_i U, U\dg \del_j U \}_{+},
 \end{array}
 \label{strain}
\eeq
where $i$ and $j$ label the space coordinates. This is a symmetric
$3\times 3$ matrix with three positive eigenvalues which we will
denote as $\lambda_a^2$, $\lambda_b^2$, $\lambda_c^2 $.
Manton~\cite{Manton} has shown that the $\lambda_i$ have a simple
geometrical interpretation.  They correspond to the length changes of
the images of any orthonormal system in a given space manifold (here
\rflat\ or \sthree) under the conformal map, $U$, onto the group
manifold, here $SU(2)\cong S_3$. Therefore the name strain tensor
which refers to such a general ``rubber-sheet'' geometry.  Among the
invariants of $K_{i j}$, three are fundamental\,\footnote{All other
invariants can be constructed from them.} and have a simple geometric
meaning:
\beq
  \begin{array}{lclcl}
   \Tr (K) &=& \lambda_a^2+\lambda_b^2 + \lambda_c^2
           &=& \sum {\rm length}^2 \\
   \half \left \{ \left ( \Tr (K) \right )^2 - \Tr(K^2) \right \}
           &=& \lambda_a^2 \lambda_b^2  + \lambda_b^2 \lambda_c^2
               + \lambda_c^2 \lambda_a^2
           &=& \sum {\rm area}^2 \\
  \det(K)  &=& \lambda_a^2 \lambda_b^2 \lambda_c^2
           &=& {\rm volume}^2 \ .
 \end{array}
 \label{invs}
\eeq
The first one measures the sum of the squared length changes of the
mapped orthonormal frame, the second one the sum of the squared area
changes and the third the squared volume change.  With the help of
Eq.\equa{strain} it is easy to see that these invariants are (modulo
normalization factors) the static energy densities of the second-order
non-linear $\sigma$ model, $(\fpi^2/2)\Tr(\del_i U \del_i U\dg)$, the
fourth-order Skyrme term, $-(\epsilon_4^2/4)\Tr[U\dg\del_i U, U\dg
\del_j U ]$, and the sixth-order term proportional to the square of
the baryon density (see Eq.\equa{L6} in section~4).  The static energy
of a general skyrmion configuration on the hypersphere has in this
language the form~\footnote{We have adopted a dimensionless form of
the Skyrme lagrangian. To obtain the dimensioned quantities, one
divides the $\lambda_i$'s and multiplies $L$ by
$2\sqrt{2}\epsilon_4/\fpi$ and multiplies $E$ by $\sqrt{2}\epsilon_4
\fpi$.}
\be
  E
    &=& \int_{S^3(L)} dV \, (\lambda_a^2 + \lambda_b^2 + \lambda_c^2
 + \lambda_b^2  \lambda_c^2 + \lambda_c^2  \lambda_a^2
 + \lambda_a^2  \lambda_b^2 ) \nonumber \\
   &=& \int_{S^3(L)} dV\, (\lambda_a-\lambda_b\lambda_c)^2 + ({\rm cycl.\
perm.'s}) + 6 \int_{S^3(L)} dV\, \lambda_a \lambda_b\lambda_c.
 \label{Egeneral}
\ee
The  last term is just $12\pi^2 B$  with the baryon (winding) number
$B$, since $\lambda_a
\lambda_b\lambda_c/(12\pi^2)$ is the Jacobian of the map
\sthree$\to$\sunit$\cong SU(2)$ - in other words the baryon (winding) number
density.  Because of the positive definiteness of the terms in the last
expression of \equa{Egeneral}\ it is obvious that any skyrmion
configuration has to respect the topological bound
\be
  E \geq 12 \pi^2 B.
\ee
Furthermore, one can immediately see that there is exactly one
possibility to satisfy the topological bound: The  insertion of
the identity map,
$\lambda_a=\lambda_b=\lambda_c$, with  $\lambda_i=1$
which
corresponds to an isometric mapping from the spacial manifold into the
target manifold and which has baryon (winding) number $B=1$.
Specifying to the case at hand where the target
manifold is $SU(2)\cong \sunit$, the unit 3-sphere, we see that the
only way for saturating the topological bound (the absolute minimum of
any skyrmion configuration with a non-zero winding number) is the
isometric mapping of the spatial
\sthree\ with $L=1$ onto the target iso-spin sphere $\sunit$.
We can therefore conclude that the topological bound can nether be
saturated by any $B>1$ system or by any skyrmion configuration in flat
space, \rflat.\footnote{Of course, we cannot exclude the possibility
that such a configuration might come infinitesimally close to the
topological bound.}\  After minimizing
$E[\lambda_a,\lambda_b,\lambda_c]$ under the constraint
\be
  \int_{S^3(L)} dV\, \lambda_a \lambda_b \lambda_c = 2 \pi^2 B
  \label{Bconstr}
\ee
Manton~\cite{Manton} and Loss~\cite{Loss} found that for $L<1$ there
is even a stronger bound on any $B=1$ skyrmion configuration,
\be
  E_{\rm identity} =6 \pi^2 \left( L + \frac{1}{L}\right ),
 \label{Esbound}
\ee
which is still satisfied by the identity map
$\lambda_a=\lambda_b=\lambda_c=1/L$ independently of the
parameterization of the $B=1$ configuration. In other words for $L\leq
1$ the identity map is the absolute minimum of any $B=1$
configuration.

Finally, it is now simple to show in full
generality~\cite{Manton,Loss} that for $L< \sqrt{2}$ the identity map
is stable for all allowed ($\delta B=0$) small-amplitude perturbations
$\lambda_i \to (1/L) + \delta_i(x)$, whereas it becomes a saddle for
$L>\sqrt{2}$.\footnote{That means that the identity map is at least a
local minimum for $1<L<\sqrt{2}$.}\ Under the small perturbation given
above the energy is given to second-order as
\be
 E[\lambda_a,\lambda_b,\lambda_c]= E_{\rm identity}
 + \left(\frac{2}{L}+\frac{4}{L^3}\right ) \int I_1
 +\left(1          +\frac{2}{L^2}\right ) \int I_2
 +\frac{4}{L^2}\int I_3 + O(\delta^3)
  \label{Esecond}
\ee
with $I_1=\delta_a +\delta_b + \delta_c$,
$I_2= \delta_a^2+\delta_b^2 +\delta_c^2$
and $I_3=\delta_a \delta_b +\delta_b \delta_c +\delta_c \delta_a$.
Using the $B$-constraint \equa{Bconstr}
we have up to third order corrections the relation
\be
  \frac{1}{L^2} \int_{S^3(L)} dV\, I_1 =
    - \frac{1}{L} \int_{S^3(L)} dV\, I_3
       + O(\delta^3).
          \nn
\ee
After inserting this relation in Eq.\equa{Esecond} and using the fact
that the $\delta_i$'s are local variations, one can finally find that
the  identity map is stable against small-amplitude fluctuations,
{\it i.e.} $E[\lambda_a,\lambda_b,\lambda_c]> E_{\rm identity}$, as
long as the following local inequality holds
\be
 \left (1 + \frac{2}{L^2} \right ) (\delta_a^2+\delta_b^2+\delta_c^2)
 - 2(\delta_a \delta_b +\delta_b \delta_c +\delta_c \delta_a ) >0,
\ee
in other words as long as $L< \sqrt{2}$. This concludes Manton's
general proof - independent of any parameterization - that the
identity map is at least a local minimum up to $L_c=\sqrt{2}$ and
becomes a saddle for $L>\sqrt{2}$.

So far we have not specified any parameterization. In the following we
will specialize to the familiar hedgehog parameterization. It has the
following properties: (a) according to a general theorem by
Palais~\cite{Palais} about reduced variational equations for symmetric
fields a variational solution of hedgehog form is a solution of the
full set of Euler-Lagrange equations of the Skyrme model, (b) the
hedgehog solution is the energetically lowest solution known so far in
the $B=1$ sector in \rflat\ and (c) it is numerically proven that the
$B=1$ hedgehog solution is at least a local minimum.  Hedgehog
configurations on the hypersphere are conveniently described in terms
of the conventional `polar' coordinates $\mu$, $\theta$, $\phi$ on
\sthree\ with $0 \leq \mu, \theta \leq \pi$ and $0 \leq \phi \leq 2
\pi$, so that a typical point on \sthree\ has Cartesian coordinates
\beq
   (L \cos \mu,\ L \sin \mu \cos \theta,\ L \sin \mu \sin \theta \cos
   \phi,\ L \sin \mu \sin \theta \sin \phi ) .
   \label{polar}
\eeq
In these coordinates the metric is $ d s^2 = L^2 ( d \mu^2 + \sin^2\mu
\,  d \theta^2+ \sin^2\mu \,\sin^2\theta\, d \phi^2) $ and the
volume element is
\beq
   d V = L^3 \sin^2\mu \,d \mu \,\sin\theta\,d \theta\, d \phi \ .
   \label{vol}
\eeq
The field components of a hedgehog of  baryon number $B$
on \sthree\ have then the form
\begin{eqnarray}
   \Phi^0  & = & \cos {f(\mu)}                         \nonumber \\
   \Phi^z  & = & \sin {f(\mu)} \cos{\theta}            \nonumber \\
   \Phi^x  & = & \sin {f(\mu)} \sin{\theta}\cos{\phi}  \nonumber \\
   \Phi^y  & = & \sin {f(\mu)} \sin{\theta}\sin{\phi}   \quad ,
   \label{hedgehog}
\end{eqnarray}
where the ``radial" profile function $f(\mu )$ is subject to the
boundary condition $f(0)=0$ and $f(\pi ) = B \pi$, $B$ integer. The
topological winding-number $B$ is of course the baryon charge of the
corresponding field configuration.  The field components, $\fpi \phia$
($\fpi$ is the pion-decay constant) should be identified with the
familiar $\sigma$, $\pi^z$, $\pi^x$, $\pi^y$ fields.  In the normal
quaternion representation the hedgehog ansatz reads
\be
 U \equiv \Phi^0 + i{\vec \tau}\cdot{\vec \Phi}= \exp(i{\vec \tau}\cdot
  \hat{r}(\theta,\phi)\, f(\mu))
  \label{uform}
\ee
where $\hat{r}(\theta,\phi)$ is the usual radial ${\bf S^2}$ unit vector.
Note that for the hedgehog configuration the eigenvalues of the strain tensor
are simply $\lambda_a^2=f'^2/L^2$, $\lambda_b^2=\lambda_c^2
=\sin^2 f/(L^2 \sin^2 \mu)$ where $f'$ stands for $\frac{df}{d\mu}$.

The energy of the hedgehog field configuration \equa{uform}\ on \sthree\
is~\footnote{Note in order to obtain the dimensioned quantities, one
multiplies $L$ in \equa{Ehedgehog} by $2\sqrt{2}\epsilon_4/\fpi$ and
$E$  by $\sqrt{2}\epsilon_4\fpi$.}
\be
    \lefteqn{ E= 4 \pi L \int_{0}^{\pi} \sin^2 {\mu} d\mu \,
    \left ( {f^\prime}^2 + 2 \frac {{\sin^2 f}}{{\sin^2 \mu}} \right ) }
     \nonumber  \\
    & + &{4 \pi }\frac{1}{L} \int_{0}^{\pi} \sin^2 {\mu} d\mu \,
       \left ( \frac{\sin^2 f}{\sin^2 \mu}
    \left [ 2 {f^\prime}^2 + \frac {{\sin^2 f}}{{\sin^2 \mu}} \right ]
    \right ).
    \label{Ehedgehog}
\ee
Variation of \equa{Ehedgehog}\ with respect to
the radial profile function $f(\mu)$ leads us to the Euler equation
\be
 \frac{d^2 f}{d \mu^2} + \frac{\sin 2\mu}{\sin^2 \mu} \frac{d f}{d\mu}
- \frac{\sin 2 f}{\sin^2 \mu}\qquad\qquad \qquad \qquad & & \nonumber \\
+\frac{2}{L^2} \frac{\sin^2 f}{\sin^2 \mu} \frac{d^2 f}{d \mu^2}
+ \frac{\sin 2 f}{L^2 \sin^2 \mu} \left \{ \left ( \frac{d f}{d \mu}
\right )^2
 -\frac{\sin^2 f}{\sin^2 \mu} \right \} &=& 0 \ .
 \label{eom}
\ee
Again because of Palais' general theorem~\cite{Palais},
solutions of (\ref{eom}) yield
solutions of the full set of Euler-Lagrange equations of the Skyrme model on
\sthree\ when substituted into \equa{hedgehog} or \equa{uform}.
It is very simple to check that one solution
of Eq.\equa{eom}\ (in the case $B=1$) is the  map $f(\mu) = \mu$
which
corresponds to the uniform mapping of \sthree\ onto
the isospin manifold ${\bf S^3}(1) \cong
SU(2)$. One can easily see that for
the uniform hedgehog map, $f(\mu)=\mu$, the eigenvalues of the strain tensor
\equa{strain}\ are simply $\lambda_a^2=\lambda_b^2=\lambda_c^2=1/L^2$. In other
words it is a special parameterization of the identity map.
The uniform map, $f(\mu)=\mu$, is a solution for all $L$, but has no
finite $L\to \infty$ limit.
The energy associated with the uniform map is of course given by
\equa{Esbound}.
As mentioned  for $1 <L < \sqrt{2}$, the uniform map is
still at least  a local minimum.
For $L > \sqrt{2}$ there exist a lower energy configuration
which represents a skyrmion localized about one point, {\it i.e.}
$f(\mu) \neq \mu$ in the hedgehog parameterization
\equa{hedgehog}.\cite{Manton,S3}\
In fact in this parameterization
the identity map bifurcates into two solutions of the
same energy concentrated around the north or south pole of the
hypersphere which are related by~\cite{S3}
\beq
 f_N (\mu)= \pi - f_S (\pi-\mu).
  \label{northsouth}
\eeq
In the limit
\ltoinf, one recovers the usual flat space skyrmion.\cite{S3}\
The identity map has $\rm O(4)$ symmetry ({\it see} Eq.\equa{hedgehog})
and is therefore  completely uniform in energy and baryon density. For
$L > \sqrt{2}$ the  symmetry  of the energetically most favorable
solution is broken to $\rm O(3)$ indicating that the solution is
localized in energy and baryon distribution. The bifurcation at
$L =\sqrt{2}$ is of the standard ``pitchfork'' type corresponding to a
second order phase transition.   In Landau's scheme the bifurcation  is
characterized by a symmetric quartic polynomial whose quadratic term
changes sign.

In order to discriminate the delocalized (high-density) phase from the
localized (low-density) phase in a way which can be generalized to
few-skyrmion systems on the hypersphere or to flat  space
array-calculations, several candidates for an order parameter
were studied  in
ref.~\cite{S3}: One candidate is the integrated squared deviation
of the local energy/baryon density from the averaged density normalized
by the squared  averaged density times the hypersphere volume. It
converges to the value 1 for the localized solution in the limit $L\to
\infty$, for smaller $L$ it decreases and shows a square-root
bifurcation at $L=\sqrt{2}$ to the value zero belonging to the identity
map. The interpretation would be that this order parameter signalled
deconfinement in the energy and baryon distribution. This is, however, a
special property of the $B=1$ system on \sthree. For few-skyrmion
systems~\cite{S3,O2O2} and flat space arrays~\cite{AndyJac,fcc} this
parameter never
becomes zero in the high-density phase, although there is still a
sizable decrease in its value from the low-density to the
high-density phase. The second suggestion of ref.~\cite{S3}
on the other hand works also in these cases. It is the squared chiral
expectation value
\beq
   \langle \sigma/\fpi \rangle^2 + \langle \vec{\pi}/ \fpi \rangle^2,
  \label{chiraldemo}
\eeq
where the $\sigma$ and pion fields are averaged over the hypersphere
volume (or the cell volume for periodic arrays).
Because of the residual $O(3)$ symmetry for the hedgehog (or
discrete symmetries in the array calculations) which cause $\langle \vec{\pi}
\rangle$ to be zero it has the same content as the parameter
$\langle \sigma \rangle$ alone. For the identity map it is obvious ({\it see}
Eq.\equa{hedgehog}) that the parameter \equa{chiraldemo}\
vanishes. In the case of the localized solution
it approaches the value 1 with increasing $L$, since in most of the
space the $U$ field of the localized skyrmion is approximately close to
the vacuum value $U_0=1$ and only markedly deviates
from this in the small region where the skyrmion is located.
The interpretation is that  a zero in this order parameter
signals chiral symmetry restoration or at least ``chiral democracy", since
there is no escape from the ``chiral-circle" constraint
$\sigma^2+ {\vec{\pi}}^2=\fpi^2$.
But nevertheless as shown for few-skyrmion systems on the
hypersphere~\cite{S3,O2O2} and for flat space arrays~\cite{AndyJac,fcc}
the $\sigma$ and pion profiles of the high-density phase are so
equally distributed about the chiral circle that their averages vanish.

Let me summarize the $B=1$ results on \sthree: There is a second-order
transition at a baryon density $1/(2\pi^2 {L_c}^3)$ for $L_c=\sqrt{2}$
from a localized and chirally broken low-density solution to a
``chirally restored" high-density solution. Restoring dimensionful
parameters into the Skyrme lagrangian (by using the empirical value of
the pion decay constant $\fpi=93$~MeV and an $\epsilon_4=0.0743$
guaranteeing a reasonable value for $g_A$) one finds
\beq
  \rho_c = \left(\frac{\fpi}{2\sqrt{2} \epsilon_4}\right )^3
\,\frac{1}{\sqrt{2}^3}\, \frac{1}{2\pi^2} \approx 0.20 {\rm fm}^{-3},
\eeq
which is far too low.\footnote{This might improve if the stabilization
term is replaced by more realistic ones. Furthermore all contributions
from kinetic terms are neglected. Finally, there is the problem of the
missing central attraction~\cite{interaction} which is probably linked
to loop corrections~\cite{softpion} or missing $1/N_c$
corrections~\cite{Nccorr}. Any additional attractive
term tends to localize the skyrmion and will therefore
move the transition point up to higher densities.}\
(remember that the nuclear matter density is
$\rho_{\rm nm} = 0.16 {\rm fm}^{-3}$). The important result, however,
is that the \sthree\ value for $\rho_c$ is more or less the same than the
one found in the periodic array calculations~\cite{Wuest,AndyJac,fcc}
under the same input parameters,
approximately $\rho_c =0.17 {\rm fm}^{-3}$. Furthermore, the
corresponding transition densities $\rho_c$ in the
few-skyrmion calculation~\cite{S3,O2O2} interpolate between these two
numbers. Thus the identification of the hypersphere formalism with the
periodic array calculation seems to be justified even quantitatively.
\newpage

\section{Indications for a Chiral Symmetry Restoration}
In this section I will present some arguments ({\it see}
refs.~\cite{chiral,WB90} for more details) why the transition reviewed in
the preceeding section should be interpreted as chiral symmetry
restoration. As mentioned the complete uniformity of the high-density
$B=1$ solution on \sthree\ is linked to the strong condition $f(\mu)=\mu$
and does not generalize to few-skyrmion systems~\cite{S3,O2O2} on
\sthree\ or periodic arrays~\cite{Wuest,AndyJac,fcc} in flat space.
However, the ``chiral democracy'' as discussed at the end of the
preceeding section can  even be achieved via a   weaker condition
\beq
  f(\mu)= \pi - f(\pi-\mu)
 \label{weak}
\eeq
which indicates that in the high-density phase there is a symmetry  about the
hypersphere equator between the ``northern" and the ``southern"
hemisphere. In fact this is just the  half-skyrmion symmetry~\cite{Goldhaber}
which is  a common signal of the high-density phase for few-skyrmion systems
on  the hypersphere~\cite{S3,O2O2} as well as for periodic
arrays in flat space~\cite{Goldhaber,variational}.
So the chiral symmetry
restoration seems to be prior to the delocalization. In fact this can be
tested by adding a (pion mass) term which explicitly breaks chiral
symmetry:\cite{chiral}
\beq
 \cL_{{\rm SB}} = \frac{{m_\pi}^2\fpi^2}{4} \Tr(U+U\dg-2) .
 \label{LSB}
\eeq
When this term is added, the phase transition is not a sharp one any longer,
but it is smeared out. The chiral order parameter \equa{chiraldemo}
still approaches zero for higher and higher densities,
but never actually becomes exactly zero. Of course, this is just a consequence
of the explicit breaking of the chiral symmetry. A second effect of adding
the term \equa{LSB}\ is a small shift of the now approximate phase transition
to higher densities, since the term induces  an attractive force.

In ref.~\cite{chiral} it is furthermore argued that the vanishing of the
order parameter $\langle \sigma \rangle$ (for the Skyrme model without
explicit symmetry breaking of course) implies  the vanishing of the
matrix element for pion decay $\langle 0| A^a_\mu (x) | \pi^b \rangle$
in case the axial current $A^a_\mu$ is calculated in the mean-field
approximation in the frame-work of a $\sigma$-model, and therefore the
vanishing of the ``effective" pion decay constant and the quark
condensate.

The most convincing arguments why the phase
transitions in the Skyrme model ought to be  identified with
a chiral symmetry restoration are the
following two: In the $B=1$ case on \sthree\ it can be shown that
the transition from the low-density phase to the
high-density phase is accompanied
\begin{romanlist}
\item
by the formation of complete chiral multiplets
in the fluctuation spectrum
\item
and by the vanishing of the three pionic Goldstone modes from the
spectrum.
\end{romanlist}

These two points are treated at length in ref.~\cite{WB90} where the
local stability of the uniform {\em and} the localized skyrmion solution on
\sthree\ and thus their small-amplitude normal modes  are investigated.
If all normal modes have positive energy, the solution is locally
stable. If some normal modes have negative energy, the solution is a
saddle point.
The study of the small-amplitude fluctuations about the $B=1$ solutions on the
hypersphere revealed the following scenario:
\begin{romanlist}
\item In general, the modes about a hedgehog solution have just
the usual $O(3)$ degeneracies, {\it i.e.} a degeneracy $2N+1$ in terms
of the principal quantum number $N\geq 1$ of the modes.
\item
The modes about the identity map on the other hand
can be classified according to the
following representations of the group $O(4)$:
the symmetric tensor representation
$(N,0)$ which has $(N+1)^2$ degenerate states and as static
eigenvalues~\footnote{We use the phrase ``static eigenvalue'' as a
short-hand notation for the change of the static energy under the
small perturbation of the fluctuation-mode.}
\beq
\lambda_{N,0} = L\left\{ N(N+2)-4\right \} + 2\left\{ N(N+2)-2\right \}/L,
 \label{nzero}
\eeq
and the
$(N,1)$ representation with a $2((N+1)^2-1)$ degeneracy and the static
eigenvalues
\beq
\lambda_{N,1} = \left\{ N(N+2)-3\right \}(L+1/L).
 \label{none}
\eeq
Thus their degeneracy exceeds by far the one of the modes about
localized hedgehog solutions. Furthermore for each mode about the
identity map with a given parity, there exists at least one degenerate
mode of opposite parity.  Using the covering group $SU(2)_L \times
SU(2)_R$ of $SO(4)$ it can be shown (see ref.~\cite{WB90}) that the
$(N,0)$ modes belong to a trajectory based on the four-fold degenerate
$(1/2,1/2)$ representations of $SU(2)_L\times SU(2)_R$, while the
$(N,1)$ modes belong to a trajectory based on the six-fold degenerate
$(1,0)+(0,1)$ representation.  Thus the modes form complete multiplets
of chiral symmetry either by the additional $(\sigma,\vec{\pi})$
degeneracy of the $(N,0)$ modes \equa{nzero}\ or by the parity doubling of
the $(N,1)$ modes \equa{none}. Note that in ref.~\cite{WB90}
the $(N,0)$ modes were identified as purely ``electric'' grand-spin modes
(but with  a degeneracy between {\it e.g.} grand spin ${\bf K}=0^{+}$ and
${\bf K}=1^{-}$ modes), while
the $(N,1)$ modes were identified as
degenerate ``magnetic'' and ``electric'' modes
of the same grand spin ${\bf K}$, but opposite parity.
The $(N=1,1)$ modes are the sixth zero modes about the identity map.
\item
At high densities ($L<\sqrt{2}$) all normal modes about the identity map
have positive energies, the identity map is stable.
There are four degenerate low-lying modes in the spectrum, the
four $(N=1,0)$ modes, one with
$\sigma$ quantum numbers and three with pion ones
indicating that the $\sigma$ and all three $\pi$-fields are degenerate and
treated on the same footing. They have the static eigenvalue
\beq
    \lambda_{N=1,0} = -L + 2/L.
\eeq
\item
The four $(N=1,0)$ modes cross zero at $L=\sqrt{2}$ and become negative for
$L>\sqrt{2}$ signalling there the instability of the identity map.
\item
At $L=\sqrt{2}$ there is the bifurcation into the normal localized
hedgehog
solution with the reduced $O(3)$ symmetry.
All modes about the localized solution  are stable for $L\geq \sqrt{2}$.
The modes with $N\geq 1$ have (with the exception
of some accidental degeneracies for special values of $L$) only the usual
$O(3)$  degeneracy $2N+1$. However, there exist 9 zero
modes.\footnote{Note that  a hedgehog in flat space has only 6 zero
modes: 3 rotational (which cannot be separated from iso-rotational ones
because of the hedgehog symmetry) and 3 translation ones.}\ The three
additional zero modes (compared with the number of zero modes for
the identity map) correspond to the three unstable pionic $(N=1,0)$
modes of the identity map. Their fourth partner, the $\sigma$-like mode,
corresponds to a positive stable excitation about the localized
hedgehog (to an infinitesimally small conformal variation in
the profile function $f(\mu) \to f(\mu)+\delta(\mu))$.
The breakdown of the continuous symmetry $O(4)$ of the identity
map at $L=\sqrt{2}$ to the  $O(3)$ symmetry
of the localized for $L>\sqrt{2}$ energetically favorable hedgehog solution
is therefore closely linked to  the appearance of three new zero modes.
\item
These three zero modes should be in fact interpreted as the three pionic
Goldstone modes linked to the spontaneous breakdown of chiral symmetry.
The arguments for this identification
are the following: Whereas six of the nine zero
modes stay normalizable even in the limit $L\to \infty$, three have no
finite $L\to\infty$ limit. This is consistent with the fact that the flat space
hedgehog has only six zero modes. Furthermore when the chiral symmetry
breaking term \equa{LSB}  is added to the Skyrme
lagrangian, the above mentioned normalizable zero modes stay zero modes,
whereas the three additional zero modes are shifted in energy by the
pion mass. These are the three pionic Goldstone modes which become
massive when an explicitly chiral symmetry breaking term is added and
which show up as non-normalizable plain wave excitations of the vacuum in
the flat space limit.
\end{romanlist}
Since the three Goldstone modes disappear for the high-density
delocalized phase and the high-density modes form complete
chiral multiplets, the chiral symmetry restoration is established.

\section{Generalizations}
\subsection{Contact Terms}
In the following we will show that the form of the Skyrme model \equa{L24}\
and especially the stabilization by the fourth-order Skyrme term
is not a necessary precondition for the very existence of the chiral phase
transition.  Let us for instance
replace the fourth-order stabilizing term in \equa{L24}\ by a
stabilizing term of sixth-order ({\it see e.g.} ref.~\cite{E2})
\beq
  \cL_{2,6} =\Frac{{f_\pi}^2}{4}\Tr \left( \partial_\mu U\dg
                                      \partial^\mu U \right )
 -c_6 B_\mu B^\mu
    \label{L6}
\eeq
(note $c_6 >0$) where $B_\mu$ is the topological baryon current
\beq
    B^\mu = \frac{\varepsilon^{\mu \nu \alpha \beta}}{24 \pi^2}
                \Tr \left( (U\dg \del_\nu U)\,
                      (U\dg \del_\alpha U)\, (U\dg \del_\beta U) \right).
   \label{barcur}
\eeq
As reported in section~2 the fourth-order Skyrme term
as well as the sixth-order term and
the second-order non-linear sigma model
term have a geometrical meaning when expressed in
the strain tensor language. One can apply therefore Manton's
general machinery~\cite{Manton}  to show that the sixth-order term
allows for the identity map as a solution and to construct the critical
value  $L_c$ where the identity map of the model \equa{L6} becomes
unstable and bifurcates into a localized solution.
In fact all qualitative  phenomena show up as  before. Naturally the value
of $L_{\rm min}$ for the  minimum of the static energy
is in general not the same for both models
and more importantly the critical hypersphere radius $L_c$
where the chiral restoration
occurs is
\beq
   L_c = (3)^\fourth L_{\rm min}
\eeq
for the model \equa{L6}
instead of  $L_c=\sqrt{2} L_{\rm min}$ as it was the case for  the
usual Skyrme model \equa{L24}.\cite{S3}

As shown in ref.~\cite{resum}\
any lagrangian which (a) is  a polynomial of the fundamental
invariants \equa{invs} of the strain tensor, (b) which  has the usual vacuum
properties and (c) which allows for (locally)
stable $B=1$ Skyrmion solutions, leads to a chirally restored phase at
high densities.
In addition the position of the phase transition can be
uniquely derived just from the lagrangian  without solving any equation
of motion. Let us discuss for this purpose a lagrangian of the analytic form
\beq
\cL_{\cG} \equiv -\cG (-\cL_2, \cL_4, \cL_6) =
          -\sum_{j_1,\, j_2,\, j_3 =0}^{\infty}
                      A_{j_1 j_2 j_3} \,(-\cL_2)^{j_1} (-\cL_4)^{j_2}
                                        (-\cL_6)^{j_3}
   \label{LG}
\eeq
where $j_1, j_2, j_3$ are integer indices and $A_{j_1 j_2 j_3}$ Taylor
coefficients.
The arguments of the analytic function $\cG$ and the signs
are chosen in such a way,
that the corresponding static energy density has the simple form
\beq
\cE \equiv \cG (\cE_2, \cE_4, \cE_6) = \sum_{j_1,\, j_2,\, j_3 =0}^{\infty}
                      A_{j_1 j_2 j_3} \,{\cE_2}^{j_1} {\cE_4}^{j_2}
                                         {\cE_6}^{j_3}.
 \label{EG}
\eeq
The function $\cal G$ should be of course positive definite to ensure a
positive energy density and ${\cal G}(0,0,0)\equiv 0$ ({\it e.g.,}
$A_{000} =0$) to ensure triviality in the $B=0$  sector. By inserting $f=0$
or $f'=0$ one furthermore learns that $\cG(x,0,0)\geq 0$ and
$\cG(x,x^2/4,0)\geq 0$ for any real $x\geq 0$. The first inequality signals
that a possible negative term expressed solely by the second-order term
can nether be compensated
by adding fourth- or sixth-order terms in {\em any}
combination to the static energy density. Therefore
Skyrme-type models which  at
fourth-order are not positive definite cannot be saved by the addition of a
sixth-order term proportional to $B_\mu B^\mu$ or any power of this term.

Note that the energy density \equa{EG}
is the most general symmetric function of the
eigenvalues, $\lambda_i^2$, of the strain tensor $K_{ij}$
\equa{strain} and includes all forms which can come form lagrangians
which involve arbitrary combinations of powers of first derivatives of
the fields to even order.

In the chirally restored (uniform) regime on the hypersphere the energy
densities of the second-, fourth- and sixth-order term, $\cE_2$, $\cE_4$
and $\cE_6$,  have the simple form (modulo a prefactor
which can be incorporated into the Taylor coefficients $A_(j_1 j_2 j_3)$)
\beq
       \cE_{2i} = \frac{3}{L^{2i}}, \qquad i=1,2,3.
\eeq
In ref.~\cite{resum} the eigenvalues of all possible small amplitude
(static) perturbations around the {\em identity} map were obtained as the
following    analytical result:
\beq
   \lambda_{N,a}^{\cG} =\left\{
               \sum_{i=1,2,3} \lambda_{i,N,a}\frac{\del}{\del \cE_{2i}}
            +\lambda_{3,N,a} \frac{2 L^2}{3}
     \left (\frac{\del}{\del \cE_2}+\frac{2}{L^2}\frac{\del}{\del \cE_4}
                +\frac{3}{L^4}\frac{\del}{\del \cE_6} \right )^2 \right \}
              {\cal G} (\cE_2, \cE_4, \cE_6)
  \label{NaG}
\eeq
with
\beq
\lambda_{i,N,a} = i \left \{ N(N+2)+a^2-4 -\frac{i-1}{2}
   \left( \{a (N+1)\}^2-4 \right )
     \right \} L^{3-2i}  .
 \label{lina}
\eeq
where the indices $N\geq 1$ and $a=0,1$ characterize the
allowed $O(4)$ representations:\footnote{{\it See}  section~3 for the
corresponding results for the usual Skyrme model \equa{L24}.}\
$(N,a=0)$, the symmetric tensor
representations  which have $(N+1)^2$ degenerate states
and the $(N,a=1)$ representations which have  a $2\{(N+1)^2-1\}$ degeneracy
({\it see} refs.~\cite{WB90,resum} for further details).
Note that in the case $a=1$ the expression for $\lambda_{N,a=1}^{\cG}$
simplifies to
\beq  \lambda_{N,a=1}^{\cG}= L \{ N(N+2)-3 \} \left ( \frac{\del}{\del
\cE_2}
   + \frac{1}{L^2} \frac{\del}{\del \cE_4} \right )
                                   {\cal G} (\cE_2,\cE_4,\cE_6) .
   \label{magev}
\eeq
A negative value for any eigenvalue $\lambda_{N,1}$
in \equa{magev} would mean that
infinitely many of the $(N,1)$ modes would pass through zero energy
{\em simultaneously}. This is probably indicative of a
non-perturbative configuration of lower dimension, {\it e.g.}, a
string-like configuration. Fortunately, one can exclude this
pathological instability, since it violates the constraint that the
model \equa{LG} should have the same vacuum properties as the
nonlinear $\sigma$ model (see ref.\cite{resum} for a discussion about
this point). So the $(N,1)$ fluctuations cannot lead to an
instability. In case the identity map is (locally) stable in the
high-density region ({\it i.e.} all $\lambda_{N,a}^{\cG}\geq 0$), the
monotonic increase of the eigenvalues $\lambda_{N,a}^{\cG}$ of the
small-amplitude normal modes with increasing $N$ is guaranteed. Then,
the critical $L_c$ where the identity map becomes unstable against
small-amplitude perturbations is given by the value of $L$ where the first
$\lambda_{N,0}^{\cG}$ in Eq.\equa{NaG} becomes negative.

Finally, let
\beq {\cG} (\cE_2,\cE_4, \cE_6) \to \cE_2 \qquad {\rm for}\
\cE_{2i}\to 0
 \label{lowdens}
\eeq
and
\beq {\cG}  (\cE_2,\cE_4, \cE_6) \to C_1 {\cE_2}^2 + C_2 {\cE_4}
    \qquad {\rm for}\ \cE_{2i}\to \infty
 \label{highdens}
\eeq
where $i$ runs from 1 to 3 and $C_1\geq 0 $ and $C_2 >0$ are fixed
positive coefficients. The condition \equa{lowdens} enforces that
asymptotically for low densities (large $L$) the lagrangian \equa{LG}\ is
dominated by the non-linear sigma model lagrangian, whereas for high
densities  (small $L$) it scales as a free Fermi gas ({\it i.e.} $\cE_\cG
\propto 1/L^4$). The low density behavior is of course indisputable, the
high density constraint on the other hand requires the
additional input that such
effective models can  be applied even for densities where the
underlying theory, QCD, becomes to leading order a free Fermi gas.

Back to the stability analysis: Note that the signs of the static
eigenvalues $\lambda_{N,0}^{\cG}$ and therefore the stability of the
uniform regime follow from the signs of the derivatives acting on
$\cG$ and from the signs of the coefficients $\lambda_{i,N,0}$
\equa{lina}. From all possible $\lambda_{i,N,a}$ with $i=1,2,3$, $N\geq
1$ and $a=0,1$ only the term $\lambda_{i=1,N=1,a=0}$ is negative,
all the other coefficients are guaranteed to
be larger than zero or at most  equal zero. (To the latter category
belong the six $O(4)$ zero modes with $N=1$ and $a=1$. Furthermore
all the terms $\lambda_{i=3,N,1}$ are zero indicating that the
sixth-order term itself has infinitely many zero modes.)\
Now taking the high-density \equa{highdens} and the low-density
behavior \equa{lowdens} into account, we can conclude that
at high densities the uniform $B=1$ skyrmion solution on the hypersphere
is bound to be stable whereas at sufficiently low densities, where the
non-linear $\sigma$ model term becomes the dominant one, the uniform
solution is unstable since $\lambda_{N=1,0}^{\cG}$ becomes negative
eventually. Furthermore we have to take into account that
in the stable regime - even
infinitesimally close to the instability - all
$\lambda_{N,0}^{\cG}$ have to increase monotonically with $N$, that
the $\lambda_{N,1}^{\cG}$ modes have to be stable by fiat and that the
eigenvalues of the modes have to be smooth functions of $L$. Then
the existence of a critical value of $L$ is
guaranteed where $\lambda_{1,0}^{\cG}$ becomes negative such that
the uniform high density phase becomes unstable.
There
is a bifurcation from the uniform solution which is chirally restored to
the usual, localized and chirally broken $B=1$ hedgehog solution. So even
for the most general {\em geometric} lagrangian \equa{LG}
built form first order field derivatives the existence of the chiral
phase transition (as discussed at length in the preceeding chapters for
the normal Skyrme model) is guaranteed under the assumption of a few
reasonable constraints (the vacuum structure, Eqs.~\equa{lowdens}
and \equa{highdens})
on the form of the lagrangian.

\subsection{Vector Meson Stabilization}
We have discussed so far lagrangians which ensured the stabilization
of skyrmions by higher-order (contact) terms in the field derivatives.
Naturally the question arises whether the existence of the chiral
phase transition is linked to the stabilization by {\em contact} terms
or whether vector mesons can be added which act over a finite range,
as {\it e.g.} $\omega$ (and $\rho$) mesons.
Let us take as an example a Skyrme-type variant  with $\omega$ meson
stabilization. (See ref.\cite{Walhoutomega} for a periodic array study
of this model.)\ The corresponding lagrangian
has the structure~\cite{omega}
\beq
  \cL_{2,\omega} = \frac{\fpi^2}{4} \Tr(\del_\mu U\dg \del^\mu U)
                  -\fourth \omega_{\mu \nu} \omega^{\mu \nu}
                  +\frac{m_\omega^2}{2} \omega_\mu \omega^\mu
                  +g_\omega \omega_\mu B^\mu
 \label{Lomega}
\eeq
where the first part is the usual non-linear $\sigma$ model term,
the second one the $\omega$ kinetic term expressed through the $\omega$
field-strength tensor $\omega_{\mu \nu}= \del_\mu \omega_\nu - \del_\nu
\omega_\mu$,
the third part is the $\omega$ mass term and the last one is the
coupling of the $\omega$ field to the topological baryon current
\equa{barcur}. As shown by Adami~\cite{Adami}\ the lagrangian \equa{Lomega}\
has stable $B=1$ solitons.

By putting an $\omega$-stabilized $B=1$ hedgehog on the hypersphere
one can easily find that one solution of the equations
of motion  is always the uniform
solution    $f(\mu)=\mu$ for the hedgehog profile and
\beq
      \omega_0= - \frac{g_\omega}{m_\omega^2} B_0
              = - \frac{g_\omega}{m_\omega^2}\, \frac{1}{2\pi^2 L^3}
     \label{ouni}
\eeq
for the $\omega_0$-component. The $\omega_0$-component is in this case
spatially constant and depends only on the hypersphere radius $L$. (As
in flat space the spatial components $\omega_i$ are identically zero
because of the static hedgehog form of the soliton profile.)\ When the
uniform solutions $f(\mu)=\mu$ and
\equa{ouni} are reinserted into the corresponding energy density
of the lagrangian \equa{Lomega}, one recovers the same structure as
for a model stabilized by a sixth-order term with $\cL_6=-
(g_\omega^2/2 m_\omega^2) B_\mu B^\mu$.  This should not come as a
surprise, since in the limit $m_\omega, g_\omega \to \infty$, under a
constant ratio $g_\omega / m_\omega$, the lagrangian \equa{Lomega}\
converges to the lagrangian \equa{L6}\ with $c_6=(g_\omega^2/2
m_\omega^2)$. Both lagrangians have the same high-density phase in the
static sector.  The physical reason for the identical behavior of the
static solutions of both lagrangians is that at high densities it is
energetically favorable that the $\omega_0$-component becomes
spatially constant, otherwise the ``costs'' in energy from a finite
spatial gradient term in this component becomes higher and higher with
increasing density. This together with the equation of motion for the
$\omega_0$-component necessarily lead to Eq.\equa{ouni} which in turn
guarantees the identical high-density behavior of both lagrangians.
However, the  $\omega$-stabilized model bifurcates at a smaller $L_c$
and the low-density behavior of
both lagrangians is different. Under increasing parameters $g_\omega$
and $m_\omega$ with the ratio $g_\omega/m_\omega$ kept constant the
critical density and the low density phase of the $\omega$ stabilized
model will approach the corresponding quantities of the
model~\equa{L6}. So the inclusion of finite range vector mesons leads to
an increase in the transition density in comparison to the corresponding
contact-term model, see also ref.\cite{Walhoutomega}.
In summary,  also  the
$\omega$-stabilized Skyrme model has a critical hypersphere
radius $L_c$ where the uniform (chirally restored) phase on the
hypersphere becomes unstable and bifurcates into a localized (chirally
broken) phase which has the usual $B=1$ hedgehog as flat space limit.

A similar behavior follows when  the $\cL_{2,\omega}$ model is extended
by the introduction of $\rho$ mesons. This can be done in different
ways. For our purposes the only essential precondition is that the total
lagrangian is still explicitly chiral symmetric. For simplicity let us
consider in the following the hidden $\rho$ meson coupling \`{a} la Bando et
al.~\cite{Bando} in a minimal way~\cite{minimal}, {\it i.e.}
\be
  \cL_{2,\omega,\rho} &=& \frac{\fpi^2}{4} \Tr(\del_\mu U\dg \del^\mu U)
  - a \frac{\fpi^2}{4} \Tr(\xi\dg\del_\mu\xi-
                \xi\del_\mu \xi\dg-2i g\rho_\mu )^2
              -\fourth \rho_{\mu \nu} \rho^{\mu \nu}
  \nonumber\\
    & &\mbox{}
            -\fourth \omega_{\mu \nu} \omega^{\mu \nu}
                  +\frac{m_\omega^2}{2} \omega_\mu \omega^\mu
                  +g_\omega \omega_\mu B^\mu +
                (\omega\rho\pi\ {\rm coupling}\ {\rm terms})
 \label{Lrho}
\ee
with $\xi\equiv \sqrt{U}$, and $\rho_{\mu \nu} =\del_\mu \rho_\nu
-\del_\nu \rho_\mu -ig[\rho_\mu,\rho_\nu]$, the non-abelian field-strength
tensor.
The second term in \equa{Lrho}
is responsible for the generation of the $\rho$ mass and
the $\rho\pi\pi$ coupling with a coupling constant $g$. (The standard
choice for the parameter $a$ is $a=2$, {\it see} ref.~\cite{Bando}.)\
Note that under the static hedgehog ansatz only the {\em spatial} $\rho_i$
components are excited. Again it is easy
to show that the equations of motion of this model are satisfied by the
uniform profiles
\be
  f(\mu)    &=& \mu \\
  \omega_0 &=& -\frac{ g_\omega}{ m_\omega^2} B_0 \\
  \rho_i    &=& \frac{1}{2 i g} ( \xi\dg \del_i \xi
                                       - \xi \del_i \xi\dg  )
\qquad {\rm with}\ \, \xi=\exp(i \tau\cdot \hat r \mu /2).
 \label{runi}
\ee
The last equation guarantees that the second term in \equa{Lrho}\
vanishes and that the $\rho$ kinetic term
behaves as fourth-order Skyrme term  with
a coefficient $\epsilon_4^2 = 1/(8 g^2)$. The lagrangian \equa{Lrho}\
(without extra $\omega \rho \pi$ couplings) has therefore the
same high density behavior as the
simple contact lagrangian
\beq
 \cL_{2,4,6}= \frac{\fpi^2}{4} \Tr(\del_\mu U\dg \del^\mu U)
             +\frac{1}{32 g^2} \Tr[U\dg \del_\mu U, U\dg \del_\nu U]^2
             - \half \left (\frac{g_\omega}{m_\omega} \right )^2 B_\mu B^\mu .
  \label{L246}
\eeq
This behavior presented here for the simplest $\omega\rho\pi$-model is
generic -- just the coefficient of the sixth-order term
in \equa{L246} may change when
$\omega\rho\pi$-coupling terms are introduced.
For all models with vector-meson stabilization which have been
studied so far~\cite{Ulfreport,Norberto,Ulfsyracuse} the following is true:
It can be numerically shown that
\begin{romanlist}
\item
the only existing $B=1$ solution at high densities is
the uniform (delocalized) hedgehog one, the static solutions are
identical to those of contact-term lagrangians with suitable
coefficients and  the static vector meson fields are equal to their
driving pionic currents ({\it see e.g.} \equa{ouni} or \equa{runi}),
\item
there exists a critical model-dependent hypersphere-radius $L_c$ where the
uniform solution becomes unstable and bifurcates into the usual localized
configuration and
\item at low densities the solutions on the
hypersphere \sthree\ converge asymptotically to the corresponding
solutions in the flat space.
\end{romanlist}
The essential conditions on these models are two-fold:
they should be explicitly chiral symmetric and
they should allow for locally stable $B=1$ hedgehog
solitons.
In case these two conditions are met, there will be necessarily a chiral
phase transition at high densities with the same features as already
discussed. Still, under realistic values for the mesonic input parameters
({\it e.g.} $f_\pi$, $\epsilon_4$, $c_6$, {\it see} ref.~\cite{E2},
or $g_\omega$ and $g$, {\it see} refs.~\cite{minimal,Ulfsyracuse,Ulfreport}))
the transition densities $\rho_c$ come out systematically too low,
$\rho_c \approx \rho_{\rm nm}$ (0.16~fm$^{-3}$). Apparently these models
lack terms which generate additional attraction.
Note that the hedgehog mass under such realistic parameters is
systematically too high ($\sim$ 1.6~GeV) whereas there is not sufficient
central attraction in the $B=2$ system.\cite{interaction}
If one would tune the
stabilizing parameters -- while keeping $f_\pi=93 {\rm MeV}$
fixed -- such that the
classical hedgehog mass would be 0.87~GeV (consistent with  a nucleon mass of
0.94~GeV), then the transition density $\rho_c$ would be approximately
three times the nuclear matter density.

\section{Second- Versus First-Order}
Whereas in the periodic array calculations second as well as first
order phase transitions were found depending on the choice of the
lattice and the twisted boundary conditions, we have so far
encountered only second order transitions between the delocalized and
the localized phase on the 3-sphere. This does not need to be the case
in general: All the Skyrme-type lagrangians presented are explicitly not
scale invariant and do not ``know" (yet) about
the trace anomaly in QCD. Let us therefore try to incorporate the same
scaling behavior as in QCD into these lagrangians ({\it see}
refs.~\cite{Schechter,Ellis,scaling}), {\it e.g.} consider the
lagrangian
\be
 \cL_{2,4,\chi}&=& \frac{\chi^2}{\chi_0^2} \,\frac{\fpi^2}{4}
 \Tr(\del_\mu U\dg \del^\mu U)
              +\frac{\epsilon_4^2}{4} \Tr[U\dg \del_\mu U, U\dg \del_\nu U]^2
   +\half (\del_\mu \chi)^2
  \nonumber\\
 & & \mbox{}
             - B_B \left (1+\left (\Frac{\chi}{\chi_0}\right)^4
          \log \left(\frac{\chi^4}{e {\chi_0}^4} \right) \right )
  \label{Lchi}
\ee
as simplest extension of the usual Skyrme lagrangian. The scalar field
$\chi$  with the vacuum expectation value $\chi_0=\langle
0|\chi|0\rangle$ is introduced with the purpose of making the first
term in
\equa{Lchi}\ scale invariant (the Skyrme term and the $\chi$-kinetic
term are already scale invariant), whereas the last term in \equa{Lchi},
the $\chi$-potential term, is adjusted to fit the trace anomaly of QCD.
$B_B$ is the ``bag constant" which can be expressed in terms of the gluon
condensate  as $B_B= (9/32)\langle 0|(\alpha_s/\pi) G^2|0\rangle$. For
values of the parameters $\chi_0$ and $B_B$ {\it see} ref.~\cite{Ellis}.
(Note that the fluctuations of the $\chi$-field correspond to
glueball-excitations.)\ In ref.~\cite{hugodang}\ it was numerically shown
that the lagrangian \equa{Lchi} again possesses  a phase transition to a
uniform delocalized solution at high densities with $f(\mu)=\mu$ and
a constant $\chi$
profile which -- depending on the choice of parameters -- can be even
$\chi=0$. Because of the insufficient accuracy in the numerics,
the authors of ref.~\cite{hugodang} missed  the fact
that the phase transition is of first order. The latter
is more or less obvious from
the form of the $\chi$-potential term which is adjusted to be minimal for
$\chi = \chi_0$. Thus it is impossible to find a smooth transition
between the constant $\chi<\chi_0$ (high-density)
profile and the localized (low density) $\chi$ profile (which
at the south pole on the 3-sphere is exactly $\chi_0$) without violating
the $\chi$ equation of motion:
\be
  \chi'' + 2\frac{\cos \mu}{\sin \mu}\chi'
  -\frac{\chi}{\chi_0^2} \frac{\fpi^2}{L^2}
  \left( f'^2+2\frac{\sin^2 f}{\sin^2 \mu} \right )
      -4 B_B \frac{\chi^3}{\chi_0^4}\log\left(\frac{\chi^4}{\chi_0^4}\right)
  = 0.
\ee
For a non-vanishing nonlinear sigma term
(as in the present case since $f(\mu)\neq 0$) the $\chi$ field cannot both be
constant and equal to $\chi_0$. So the $\chi$ field can only do the transition
from the low- to the high-density phase and vice versa by a jump.
Thus by incorporating a scalar
field \`{a} la Schechter~\cite{Schechter} in  the
Skyrme model (or its extensions) the second order phase transition can be
easily changed to a first order one. {\it See} also ref.~\cite{chiral} for a
different mechanism to achieve the same thing.

It is rather easy to extend the lagrangian \equa{LG}\ to the most
general with the QCD trace anomaly consistent geometrical
form involving only first derivatives to even order:
\be
  \cL_{\cG,\chi} &=& -\left(\frac{\chi}{\chi_0}\right)^4\,
\cG\left(      -({\frac{\chi_0} {\chi}})^2\cL_2,
               -({\frac{\chi_0} {\chi}})^4\cL_4,
               -({\frac{\chi_0} {\chi}})^6\cL_6\right ) \nonumber\\
 & & \mbox{} +\half (\del_\mu \chi)^2
             - B_B \left (1+\left(\Frac{\chi}{\chi_0} \right )^4
          \log \left(\frac{\chi^4}{e {\chi_0}^4} \right) \right ) .
  \label{LGchi}
\ee
Note that the restoration of scale invariance, $\chi=0$, is only
consistent with the lagrangian \equa{LGchi}\ if $\cG$ has the asymptotics
\equa{highdens}, {\it i.e.}
\beq
 \left(\frac{\chi}{\chi_0}\right)^4\,
\cG\left(      -({\frac{\chi_0} {\chi}})^2\cL_2,
               -({\frac{\chi_0} {\chi}})^4\cL_4,
               -({\frac{\chi_0} {\chi}})^6\cL_6\right )
   \to
   C_1 (-{\cL_2})^2 +C_2 (-\cL_4) \quad {\rm if\ } \chi \to 0.
  \label{xhigh}
\eeq
We know from what was said before that one can expect in this case a
first-order phase transition to a chirally restored phase. However,
the $\chi$-profile in the high-density phase does not need to become
zero, it can equally well be just a constant $\chi < \chi_0$.  Scale
invariance implies chiral restoration, but not vice versa.
Nevertheless, just the possibility to have a scale invariant limit at
high densities leads to strong constraints on the form of the
effective model.  If a lagrangian is considered which includes {\em
any} term higher than fourth-order in the derivatives of the pion
fields, terms of {\em all orders} have to be included as well both to
satisfy the high-density asymptotic behavior
\equa{xhigh}\ and to ensure the stability of the chiral symmetric high-density
phase.\cite{resum}\ The necessity for infinitely many terms is
consistent with the large $N_c$ philosophy.\cite{Hooft,Witten}\
Because of asymptotic freedom, the effective model which follows from
QCD to leading order in the $1/N_c$ expansion must involve infinitely
many mesons. The hope is -- as shown for the $\omega$ and $\rho$ meson
case -- that the high density phase of this infinite tower of meson
resonances can be approximated by infinitely many suitable contact
terms.  This might lead to ``dreams" of Regge trajectories.

\section{Discussion}
Even effective models of the Skyrme class (which do not possess any
quark degrees of freedom) can have a phase transitions to a
high-density phase with restored chiral symmetry.  There are
essentially just two conditions on such a model: It must have the
relevant symmetries (especially the lagrangian must be explicitly
chirally invariant) {\em and} it must treat the baryon structure and
interaction on the same footing. Normally, the phase transitions are
of second order, but one can easily change them to first-order ones by
incorporating the QCD trace anomaly for instance.  This allows also -
depending on the input parameters - for a scale invariant high density
limit. Scale invariance implies  chiral symmetry restoration, but
not vice versa.

Under the same parameter input (from the meson sector) the
periodic array calculations as well as the 3-sphere calculations give
approximately the same values for the transition densities. The
predictions,  however, are for both systematically too low, $\rho_c \approx
\rho_{\rm nm}$.  This problem may be connected with two other
deficiencies which plague the Skyrme-like models: the missing central
attraction in the $B=2$ channel and the predicted high value for the
hedgehog mass. Quantum (higher loop or higher $1/N_c$) corrections may
have a chance to cure all three problems at the same time by providing
extra non-local terms which generate attraction.

\nonumsection{Acknowledgements}

The author would like to thank A.D.~Jackson, L.~Castillejo and
C.~Weiss  for useful discussions.

\nonumsection{References}

\end{document}